\documentclass[aps,prl,amsmath,amssymb,reprint,superscriptaddress,twocolumn,showpacs]{revtex4-2}

 \usepackage{graphicx}
 \usepackage{amsmath}
 \usepackage{amsfonts}
 \usepackage{amssymb}
 \usepackage{pstricks}
 \usepackage{psfrag}
 \usepackage{epsfig}
 \usepackage{changes}
 \usepackage[ansinew]{inputenc}
\usepackage{mathptmx}
\usepackage{helvet}
\usepackage{textcomp}
\usepackage{ulem}
 \usepackage{bm}

\begin{document}


\title{ Collective Flows Drive Cavitation in Spinner Monolayers}



\author{Zaiyi Shen}
\email[]{zaiyi.shen@pku.edu.cn}
\affiliation{Univ. Bordeaux, CNRS, LOMA, UMR 5798, F-33400 Talence, France}
\affiliation{State Key Laboratory for Turbulence and Complex Systems, Department of Mechanics and Engineering Science, College of Engineering, Peking University, Beijing 100871, China}
\author{Juho S. Lintuvuori}
\email[]{juho.lintuvuori@u-bordeaux.fr}
\affiliation{Univ. Bordeaux, CNRS, LOMA, UMR 5798, F-33400 Talence, France}


\date{\today}

\begin{abstract}
Hydrodynamic interactions can give rise to a collective{ \color{black} motion} of rotating particles. This, in turn, can lead to coherent fluid flows. Using large scale hydrodynamic simulations, we study the coupling between these two in spinner monolayers {\color{black} at weakly inertial regime.} 
We observe an instability, where the initially uniform particle layer separates into particle void and particle rich areas. The particle void region corresponds to a fluid vortex, and it is driven by a surrounding spinner edge current. We show that the instability originates from a hydrodynamic lift force between the particle and fluid flows. {\color{black} The cavitation can be tuned by the {\color{black} strength of the} collective flows.  It is suppressed when the spinners are confined by a no-slip surface, and {\color{black} multiple cavity and oscillating cavity} states are observed when the particle concentration is reduced.}
\end{abstract}

\pacs{}

\maketitle


\paragraph{Introduction.---}
Collective order appears in active systems when the individual constituents change their own behaviour due to the influence from the others~\cite{vicsek2012collective}. Fluid mediated interactions are widely existent, in artificial systems, such as self-propelling colloidal suspensions~\cite{martinez2018advances,driscoll2019leveraging}, and in the living world such as bacterial baths~\cite{lauga2016bacterial,koch2011collective,mathijssen2018nutrient,dhar2022self}. In both cases they play a crucial role in the emergence of collective phenomena{\color{black}~\cite{driscoll2017unstable,bricard2013emergence,thutupalli2018flow,han2020reconfigurable,riedel2005self,yu2018ultra,aragones2016elasticity,yeo2015collective,drescher2009dancing,petroff2015fast,shen2019hydrodynamic,koizumi2020control,thijssen2021submersed}}.
Interestingly, hydrodynamic interactions are often non-reciprocal, which can have profound effects on the formation of out of equilibrium states~\cite{poncet2022soft,fruchart2021non}. For example, in a pair of rotating particles, the spinners experience a transverse force due to the advection created by the flow field of the other~\cite{lenz2003membranes,fily2012cooperative,yeo2015collective}.  At vanishing Reynolds number, these interactions have been predicted to give rise to hyperuniform states~\cite{oppenheimer2022hyperuniformity} and lead to fast crystallisation when coupled with steric repulsion~\cite{oppenheimer2019rotating}, in open systems.
When spinner crystals are confined by a solid boundary, the transverse forces have been observed to give rise to motile dislocations in odd crystals~\cite{bililign2022motile}.

\begin{figure*}
\centering
\includegraphics[width=1\textwidth]{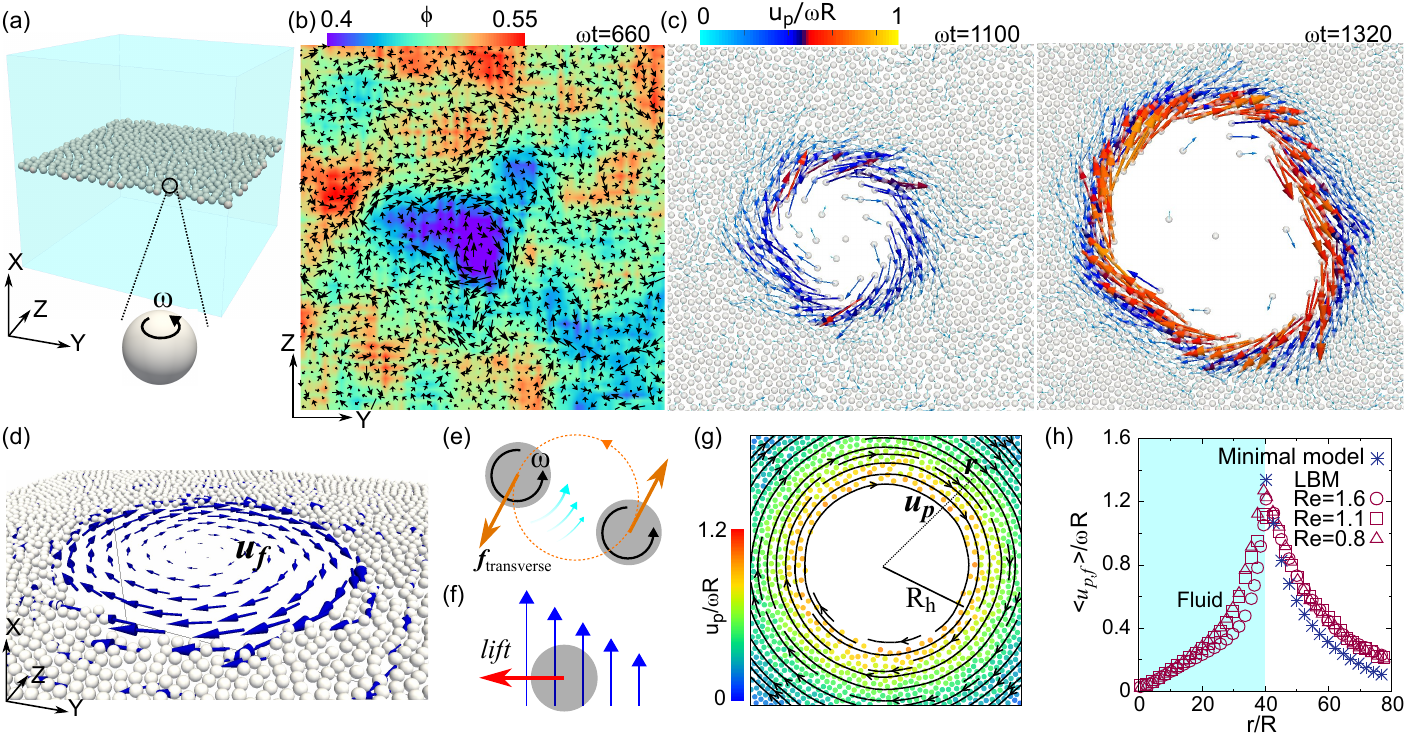}
\caption{(a) A schematic of the system. The particles are {\color{black} in a monolayer} bounded by a fluid domain and rotate perpendicular to the plane. (b,c) Top view snapshots from a simulation with $\mathrm{Re} \approx 5$. (b) At early time the system exhibits a dynamically uniform state accompanied by local concentration fluctuations leading to particle movement. (c) At later times a coherent particle current is formed. This triggers an instability of the density distribution, leading to a spontaneous formation of a particle cavity, surrounded by a spinner edge flow.
(d) At the steady state a vortex flow of the fluid is observed in the hole region.
(e) A schematic showing the transverse pair interaction {\color{black} (orange arrows) arising from the rotational flow fields.} (f) {\color{black} A Saffman type lift force resulting} from the velocity lag between the particle and imposed shear flow.
(g) A minimal model employing {\color{black} only the transverse interactions, reproduces {\color{black} the particle velocities for artificially imposed cavity.}}
(h) {\color{black} The distance $r$ dependence of the circularly averaged fluid $u_f(r)$ and particle $u_p(r)$ velocities for the hole (blue) and particle (white) regions. The circles, {\color{black} squares and triangles correspond to a full hydrodynamic simulation with $\mathrm{Re} \approx 1.6,~1.1$ and $0.8$, respectively.}} \label{hydro}}
\end{figure*}

Increasing the Reynolds number ($\mathrm{Re}$), the interactions can include inertial forces~\cite{climent2007dynamic,steimel2016emergent,aragones2016elasticity}.
 Rotating disks on a gas-liquid interface at finite Reynolds numbers have been observed to form hexagonal structures due to an interplay between a hydrodynamic repulsive force and a magnetic attraction ~\cite{grzybowski2000dynamic}. Simulations have predicted the formation of spinner aggregates arising from inertial hydrodynamic attractions~\cite{goto2015purely,shen2020hydrodynamic}.
A two-phase crystallization has been predicted in spinner monolayers bounded by a solid surface when inertial forces are included~\cite{shen2020two}.


{\color{black} Typically, in a suspension of active particles at $\mathrm{Re}\approx 0$, the hydrodynamic force on particle $i$ arising from the flow fields $\mathbf{u}_j(\mathbf{x}_i)$ of all the other particles,  can be evaluated from $\mathbf{F}_i(\mathbf{x}_i) \sim \sum_{j\neq i}\mathbf{u}_j(\mathbf{x}_i)$. 
When $\mathrm{Re}\sim 1$,  an additional force $\mathbf{F}(\mathbf{v}_i,\sum_{j\neq i} \mathbf{u}_j(\mathbf{x}_i))$ can appear, which depends on the velocity lag  $U$ between the particle velocity $\mathbf{v}_i$ and collective, activity induced, fluid flow. 
This, {\it active lift-force}, may lead to new collective phenomena in (weakly) inertial active matter.} 

In this letter, using hydrodynamic simulations, we  study this coupling between self-generated fluid flows and particle dynamics. Motivated by experimental realisations of torque driven particles trapped at interfaces~\cite{han2020reconfigurable,han2020emergence,grzybowski2002directed,grzybowski2000dynamic,grzybowski2001dynamic}, {\color{black} we simulate spinner monolayers bounded by a bulk fluid [Fig.~\ref{hydro} (a)].} We observe an instability, where the initially uniform state spontaneously separates into a particle void region surrounded by a whirling particle edge flow [Fig.~\ref{hydro} (c)]. {\color{black} This arises from a dynamic feedback-loop between the particle dynamics and the self-induced flow fields.} 
The particle void region corresponds to a fluid vortex  {\color{black} driven by the particles [Fig.~\ref{hydro} (d)]}. The onset of the instability is caused by {\color{black}the vortex flow induced} lift force acting perpendicular to the particle velocity. At the steady state cavitation, this is balanced by steric interactions between the particles in the dense phase.

\paragraph{Methods.---} The simulations are carried out by a lattice Boltzmann method (LBM), which is used to solve the quasi-incompressible Navier-Stokes equation for the fluid flow ~\cite{succi2001lattice,chen1998lattice,kevinstratford_2022_7075865,SM}. The bounce back on links method ~\cite{ladd1,ladd2} for a moving boundary~\cite{nguyen2002lubrication}, is applied to take into account the fluid-solid interactions, leading to no-slip boundary condition on the particle surface. We consider a density-matched suspension $\rho = \rho_{fluid} = \rho_{particle}=1$, and set the LBM lattice spacing $\Delta x = 1$ and time unit $\Delta t = 1$.
The spinner monolayer consists of $N$ spherical particles with radius $R = 4.1 \Delta x$. The {\color{black} particles} are driven by a torque $T$, leading to a rotational motion (around $X$) with a frequency $\omega$ [Fig.~\ref{hydro} (a)].
The spinner monolayer is placed in $YZ$ plane in a three-dimensional periodic box.
The dynamical state of the system is characterized by the rotational particle Reynolds number $\mathrm{Re}=\rho \omega R^2/\mu$, where $\mu$ is the dynamic viscosity of the fluid. We consider a weakly inertial regime, with $\mathrm{Re}$ ranging from 0.02 to 5. The particle concentration $\phi_0$ is defined as the area fraction of the monolayer $\phi_0 = N \pi R^2 / L_Y L_Z$, where $L_{X|Y|Z}$ are the simulation box lengths.  A rectangular box with $L_X=60R, L_Y=160R, L_Z=160R$ is used. {\color{black} All the simulations have $N=4000$ particles ($\phi_0 \approx 50 \%$) except for Fig.~\ref{concen}, where $\phi_0$ is varied}. A simulation using a cubic box is also performed to demonstrate that the vertical size $L_X=60R$ is large enough to minimize the periodic effects (see Fig.S1 in ~\cite{SM}) .

\paragraph{Cavitation in a spinner monolayer.---}

Considering random initial particle positions within the spinner monolayer, {\color{black} the hydrodynamic interactions between the spinners~\cite{bililign2022motile,shen2020two}, coupled with local concentration fluctuations} lead to creeping particle currents [Fig.~\ref{hydro} (b)]. 
We find that this {\color{black}quasi-}uniform state is not always stable. {\color{black} The coupling between the particle velocity and local density} can promote a positive feedback where {\color{black} the density can fluctuations grow} [Fig.~\ref{hydro} (c) and Movie S1 in ~\cite{SM}]. {\color{black} Eventually,  at the steady state, a cavitation is observed. A particle void phase is surrounded by a particle rich phase (Movie S2 in ~\cite{SM}). 
The interface is stabilised by a spinner edge current,} where the counterclockwise rotating particles form a clockwise flowing current.

{\color{black} The translational motion originates from the hydrodynamic fields created by the other particles.} 
The {\color{black} rotational flows, lead to {\color{black} interactions between the particles. At zero $\mathrm{Re}$-limit these can be captured by so called transverse forces between the spinners $f_{\mathrm{transverse}}\sim \mu\omega$ [orange arrows in Fig.~\ref{hydro} (e)]~\cite{bililign2022motile,shen2020two}.} {\color{red} {\color{black} We hypothesize that these drive the particle edge currents}}. 
{\color{black} To test this, we consider a minimal dry model where time evolution of the position $\bm{x}_i$ for the particle $i$ is governed by the {\color{black} advection arising from the flow fields of} all the other particles} 
\begin{equation}
\frac{\mathrm{d} \bm{x}_i}{\mathrm{d} t}=\sum_{j} \bm{\omega}_j \times \frac{R^{3}}{|\bm{x}_{i}-\bm{x}_{j}|^{3}}(\bm{x}_i-\bm{x}_j).
\label{min1}
\end{equation}

Starting from a state with a cavity $R_h \approx 40R$ the minimal model simulations (Eq.~\ref{min1})  reproduce the edge current  [Fig.~\ref{hydro} (g)]. {\color{black} This agrees with experiments, where edge currents have also been observed at the surface of spinner clusters~\cite{soni2019odd} and at grain boundaries~\cite{lobmeyer2022grain} at the $\mathrm{Re}\approx 0$ limit.}
The measured radial particle velocity distribution shows a very good agreement with the hydrodynamic simulation [Fig.~\ref{hydro} (h)].
This suggests that the hydrodynamic transverse forces are the driving force behind the stable edge currents observed in the full hydrodynamic simulations.
However, in the {\color{black} dry} simulations, the {\color{black} cavity is not stable.}  The hole shrinks, and finally disappears (Movie S3 in ~\cite{SM}). This implies that there is another force leading to the cavitation. We attribute this to a {\color{black} radial force $f_c=f(\mathbf{v}_i,\sum_j\mathbf{u}_j)$} arising from the coupling {\color{black} between particle edge flow and the} fluid vortex [Fig.~\ref{hydro} (d) and (f)].

\begin{figure}
\centering
\includegraphics[width=0.93\columnwidth]{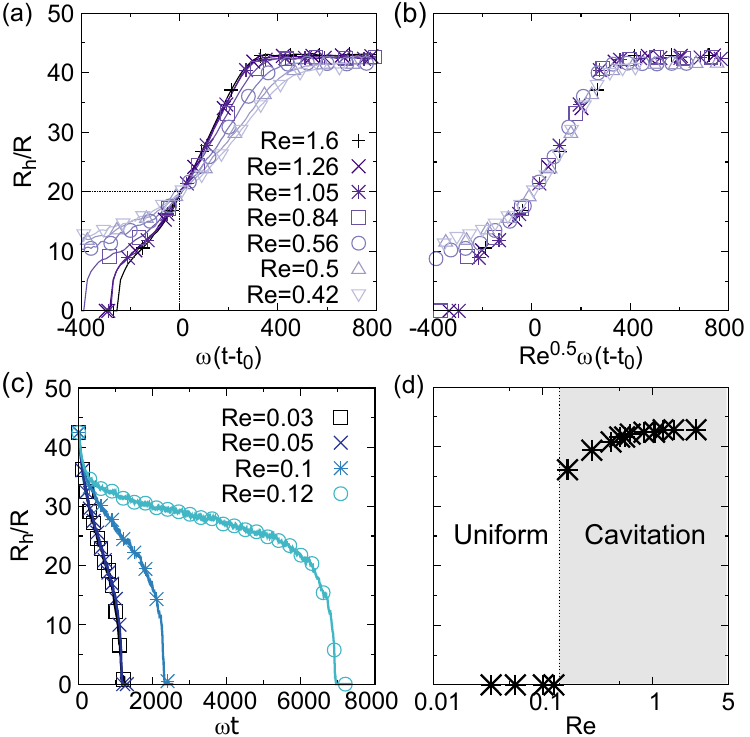}
\caption{(a) Time evolution of the hole radius $R_h$ in the cavitation state for different $\mathrm{Re}$. The hole grows and reaches a steady state. (b) {\color{black} The $R_h(t)$ data collapse when rescaled by $\mathrm{Re}^{0.5}$.} 
(c) The $R_h(t)$ for shrinking cavities when $\mathrm{Re}$ is reduced. (d) A  {\color{black} state diagram as a function of $\mathrm{Re}$. Stable cavitation  is observed for $\mathrm{Re}\gtrsim 0.2$.}
\label{inertia}}
\end{figure}

\paragraph{Collective hydrodynamic force behind the cavitation.---}
{\color{black} To study the collective dynamics,} we carry out simulations where the spinners are driven by a constant frequency $\omega$ and initially seed a small hole with a radius $R_h \approx 5R$ in the  monolayer {\color{black} (Fig.~\ref{inertia})}.}
The dynamics of hole growth is studied by measuring the effective size of the cavity $R_h(t)=\sqrt{A_h/\pi}$, where $A_h$ is the area of the particle void region (where the particle concentration $\phi < 0.8 \phi_0$). After an initial time $t_0$ {\color{black} when the hole has reached $R_h \approx 20R$}, we identify a steady growth stage $R_h = \int v_{h} dt$ {\color{black}{\color{black}~[Fig.~\ref{inertia} (a)].} The growth rate $v_{h}$ arises from the balance between the {\color{black} radial force} and a fluid drag {\color{black} $v_{h} \sim f_c/\mu$ and} the expansion continues until the particle-particle excluded volume interactions arrest the growth. 

We propose that the dominating interactions arise from the coupling between the particle dynamics and the vortex flow, similar to a Saffman lift force ~\cite{saffman1965lift} {\color{black} where a particle in a shear flow has a lift force along the shear gradient [Fig.~\ref{hydro} (f)]}.  The Saffman lift scales as  $f_{lift} \sim \mu^{0.5} U \dot{\gamma}^{0.5}$ ~\cite{saffman1965lift}, where $U$ is the relative velocity between the particle and the imposed flow and $\dot{\gamma}$ is the shear gradient. The coherent vortex flow and the translational particle motion both originate from the rotation of the particles. Thus the magnitude of $U$ and $\dot{\gamma}$ can be approximately given by $\omega$. This gives  ${\color{black}f_c} \sim f_{lift} \sim \mu^{0.5} \omega^{1.5}$, {\color{black} and the growth rate} $v_{h} \sim {\color{black}f_c}/\mu \sim \mathrm{Re}^{0.5} \omega$. {\color{black} We observe a good collapse of the {\color{black} data} with} $\mathrm{Re}^{0.5} \omega$ [Fig.~\ref{inertia} (b)]. This supports the hypothesis of the shear induced lift force driving the cavitation.

The {\color{black} lift} forces require a finite $\mathrm{Re}$. {\color{black} When $\mathrm{Re} \lesssim 0.2$  no hole formation is observed [Fig.~\ref{inertia} (d)]. Starting from the cavitation state, particle rotation induced mixing $f_m$ overcomes the inertia lift force, and the hole shrinks until} a uniform state is observed [Fig.~\ref{inertia} (c)], {\color{black} in agreement with experiments of colloidal spinners at vanishing Re~\cite{soni2019odd} and the dry simulations (Movie S3 in~\cite{SM}).}  When $\mathrm{Re}  \ll 1$, the {\color{black} closing} dynamics is dominated by {\color{black} mixing ($f_m \sim \mu\omega \gg {\color{black}f_c}\sim \mu\mathrm{Re}^{0.5}\omega$) arising from the transverse forces between the spinners.} {\color{black} The hole shrinking rate is expected to scale as $v_h \sim f_m/\mu \sim \omega$ and a collapse of the data is observed for $\mathrm{Re} \approx 0.03$ and $\mathrm{Re} \approx 0.05$ [Fig.~\ref{inertia} (c)].} {\color{black} For larger $\mathrm{Re}$, $f_m\gtrsim {\color{black} f_c}$ and the closing dynamics slows down [Fig.~\ref{inertia} (c) $\mathrm{Re}\approx 0.1$ and $\mathrm{Re}\approx 0.12$].}

\begin{figure}
\centering
\includegraphics[width=1\columnwidth]{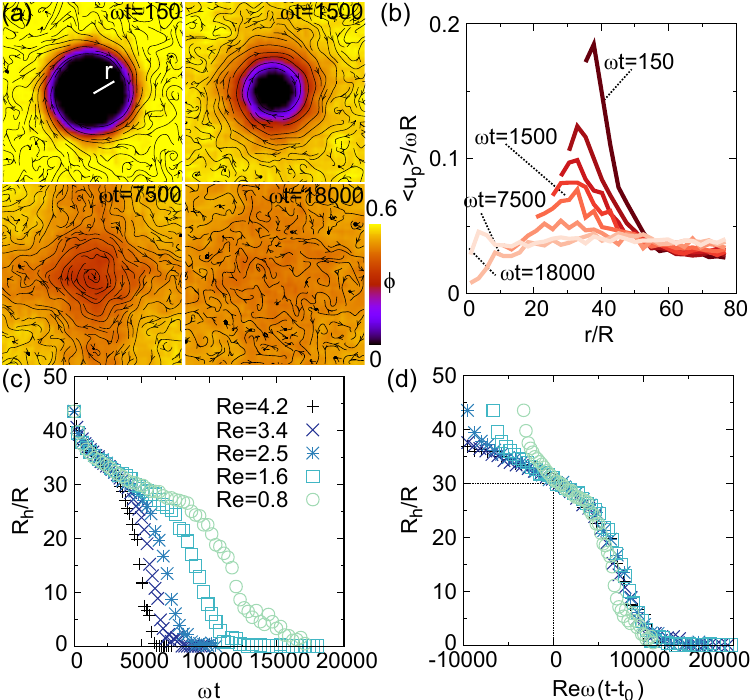}
\caption{(a) {\color{black} Snapshots of the disappearing cavity in a monolayer near a no-slip wall ($\mathrm{Re} \approx 1.6$). The streamlines correspond to the particle velocities}. (b) {\color{black} The average particle velocity $<u_p>$ as a function of the radial distance $r$.} 
The decreasing hue corresponds to increase in time. (c) Time evolution of the hole radius $R_h$ for different $\mathrm{Re}$. (d) The data for $R_h(t)$ collapse when rescaled by $\mathrm{Re}$. 
\label{wall}}
\end{figure}

 \paragraph{Confinement effects.---}
{\color{black}{\color{black} Stable cavitation} requires {\color{black} large enough} collective particle velocity {\color{black} and resulting } $\dot{\gamma}$ to ensure a sufficient lift force. {\color{black} Boundaries, such as hydrodynamic screening from surfaces can have drastic effects on the magnitude of the flow fields.}
To study this, the spinner monolayer was placed above a flat no-slip wall. Starting from an initial state with a cavity $R_h \approx 40R$, the $R_h$ is observed to decrease [Fig.~\ref{wall} (c)].
Eventually, the cavity disappears, and at the steady state only local density fluctuations are observed [Fig.~\ref{wall} (a) and Movie S5 in ~\cite{SM}].
Keeping the rotational $\mathrm{Re}$ constant, the particle velocities $u_p$ are {\color{black} observed to reduce when compared to a bulk sample. For a $\mathrm{Re}\approx 1.6$ a maximum particle velocity  $u_p\approx 0.2 \omega R$ is observed [Fig.~\ref{wall} (b)], which is much smaller than what is measured in the bulk sample} $u_p\approx1.2 \omega R$ [Fig.~\ref{hydro} (h)]. {\color{black} Thus the shear gradient $\dot{\gamma}$ and the resulting lift force are reduced.}

\begin{figure}
\centering
\includegraphics[width=1\columnwidth]{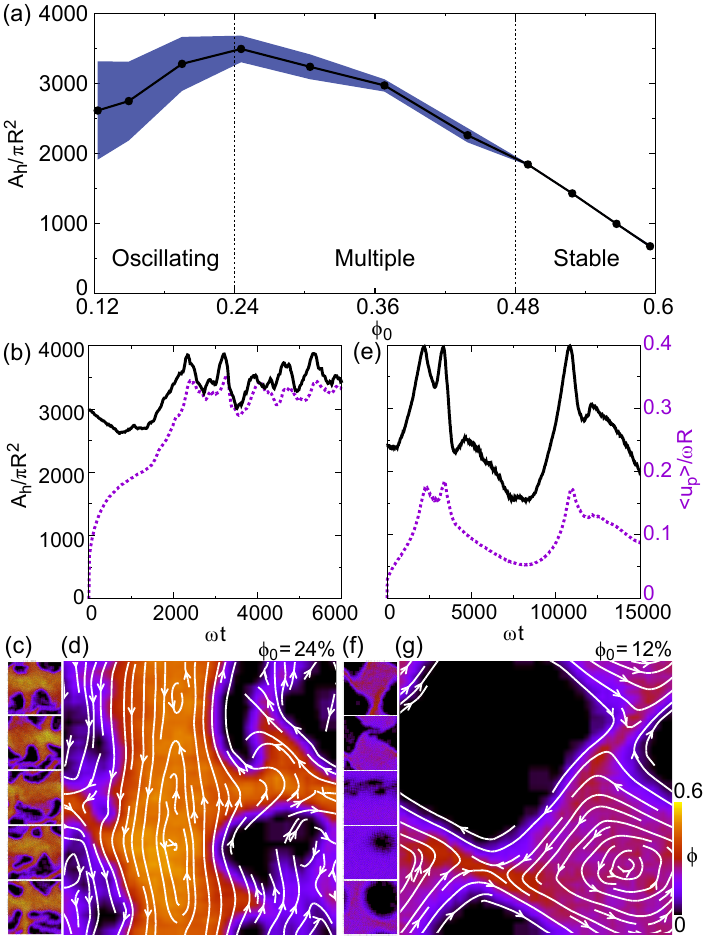}
\caption{(a) The area $A_h$ of the hole regions as a function of the overall particle concentration $\phi_0$. The blue shadow shows a standard error, {\color{black} describing the time fluctuations of $A_h$. }{\color{black} Stable, multiple and oscillating cavities are observed when $\phi_0$ is decreased}. (b-d) {\color{black} An example of the observed multiple cavity state}: $N=2000$ particles ($\phi_0 \approx 24\%$). (b) Time evolution of hole size and mean particle speed. (c) Time series snapshots of the multiple cavity state. (d) Typical local concentration and streamlines of particle velocity. (e-g) {\color{black} Oscillating cavity state}: $N=1000$ particles ($\phi_0 \approx 12\%$). (e) Time evolution of the hole size and mean particle speed. (f) Time series snapshots of the oscillating state. (g) Typical local concentration and streamlines of particle velocity. All the simulations are carried out using $\mathrm{Re} \approx 1.6$. \label{concen}}
\end{figure}

{\color{black} The inertial hydrodynamic forces can also include (pair-wise) repulsive interactions~\cite{shen2020two,climent2007dynamic}.} At finite rotational Reynolds numbers, a single particle flow field has also polar and radial components~\cite{bickley1938lxv,climent2007dynamic,grzybowski2000dynamic}. The latter corresponds to an advection away from the particle at the equatorial region, and gives rise to an effective repulsion between the spinners $f_{repulsion} \sim \mu \omega \mathrm{Re} $~\cite{shen2020two,climent2007dynamic}. {\color{black} This would lead to a reduction of the cavity with a predicted rate $v_h \sim f_{repulsion}/\mu \sim \mathrm{Re} \omega$.}  
{\color{black} In agreement with this,  after an initial transient $t_0$ when $R_h \approx 30R$,  the $R_h(t)$ data collapses  when scaled with $\mathrm{Re} \omega$ [Fig.~\ref{wall} (d)]}}, {\color{black} suggesting that the secondary flow induced repulsion dominates the near wall dynamics in agreement with previous simulations~\cite{shen2020two}.}

\paragraph{Concentration effects.---}

{\color{black} {\color{black} The overall area fraction $\phi_0$ defines the average particle separation, and thus affects the strength of the collective particle currents and the resulting fluid flow. These, in turn, determine the strength of the lift force, and thus influence the dynamics of the phase separation {\color{black} [Fig.~\ref{concen}]}.

At high particle concentrations ($\phi_0 \gtrsim 48\%$) [Fig.~\ref{concen} (a)], a stable circular hole state is observed. The hole size decreases with increasing particle concentration. When the overall area fraction is reduced ($24\% \lesssim \phi_0 \lesssim 48\%$), the circular cavity is no longer stable {\color{black} and spontaneous elongation is observed (see {\it e.g.} Movie S6 in \cite{SM}). This can lead to an imbalance of the hydrodynamic stresses at the interface, and destabilise the cavity.  Now a dynamic state is observed}, where multiple cavities can coexist simultaneously [Fig.~\ref{concen} (c) and Movie S6 in \cite{SM}]. When concentration is further lowered ($\phi_0 \lesssim 24\%$), the multiple cavity state becomes unstable. Now, an oscillating state is observed, where the hole region forms and disappears periodically [Fig.~\ref{concen} (f) and Movie S7 in~\cite{SM}]. The time evolution of the mean particle speed $\langle u_p\rangle$ and {\color{black} the cavity area $A_h$ show strong correlation [Fig.~\ref{concen} (b) and (e)].} This further {\color{black} highlights the coupling between the particle organisation and the collective flows.} 


\paragraph{Conclusions.---}
Using numerics, we have studied the coupling between the particle dynamics and the self-induced fluid flows. Considering a simple system of torque driven spinners, we have observed a spontaneous cavitation in spinner monolayers.  
At the steady state, a particle deprived region surrounded by dense spinner phase is formed. The cavity region corresponds to fluid vortex with opposite handedness than the spinners.  The vortex is driven by a steady spinner edge current at the interface between the two regions. 
The phase separation is stabilised by a lift force arising from the coupling between {\color{black}  the particle dynamics and the collectively generated fluid flow. This is crucially different from typical active matter examples, such as  motility induced phase separation and bubble formation in active Brownian models~\cite{stenhammar2014phase,shi2020self,caporusso2020motility}, which are stabilised by a density dependent swimming speed, or vortex formation in subcritical Quincke rollers, which is likely dominated by an interplay between hydrodynamic and thermodynamic forces as well as activation events~\cite{liu2021activity}.}


Our result  should be observable, for example, in experiments of magnetically rotated particles trapped at interfaces ~\cite{han2020emergence,zhang2020reconfigurable,grzybowski2002directed,grzybowski2000dynamic,grzybowski2001dynamic}, {\color{black} or by considering chiral colloidal fluids~\cite{soni2019odd}, where hydrodynamics interactions are likely dominant.  The $\mathrm{Re}\sim 1$ regime could be reached by increasing either the particle size or the rotational frequency of the  magnetic drive --- a 100$\mu$m particle spinning with 100Hz will give $\mathrm{Re}\approx 1$ in water. Further, we anticipate that the inertial term (active lift-force) could give rise to new collective dynamical states in weakly inertial active matter,  also beyond the spinner example considered here.} 

\begin{acknowledgments}
ZS and JSL acknowledge IdEx (Initiative d'Excellence) Bordeaux and the  French  National  Research  Agency  through  Contract No. ANR-19-CE06-0012 for funding, Curta cluster for computational time.
\end{acknowledgments}

\bibliography{ref}



\end{document}